# Observation of Moiré Flat Bands in Sonic Crystals

Tingzhi Liu,[†] Xingjian Zhang,[†] Qicheng Zhang, Xiying Fan, Fengcheng Wu, and Chunyin Qiu[*]

Key Laboratory of Artificial Micro- and Nano-Structures of Ministry of Education
and School of Physics and Technology
Wuhan University, Wuhan 430072, China
[†] These authors contributed equally.
[*] To whom correspondence should be addressed: cyqiu@whu.edu.cn

*Abstract.*—Recently, artificial moiré superlattices of classical waves have aroused tremendous interest, inspired by the newly emergent twistronics that focuses on the peculiar electronic properties induced by flat bands. However, so far, the moiré flat bands have not been observed directly. Here, we report the first experimental observation of ultraflat bands in moiré sonic crystals with frequency-momentum spectra, together with a real-space visualization of the hallmark localized states. Strikingly different from the established magic angle mechanism for twisted bilayer graphene, the flat bands are formed by weakly coupled moiré potential-well states inside a wide band gap, and can be realized over a broad range of twist angles. The average group velocity of the moiré localized states, as a faithful reflection of the band flatness, decays exponentially with the moiré period of the acoustic structure. Our findings, applicable to all artificial crystals, enable new possibilities for manipulating classical waves with moiré structures.

*Introduction.*—Moiré patterns can be formed by stacking two monolayers of two-dimensional (2D) materials and rotating one of them by an angle. The twist angle serves as an extra degree of freedom to manipulate electronic wavefunctions and energy bands through interlayer hybridizations. This results in numerous exotic electronic properties and optical responses [1-23] unavailable in individual monolayers. Typical examples include: unconventional superconductivity [1-6], moiré-trapped excitons [7-9], emergent magnetisms [10,11], topological states [12-15], etc. Many of the strongly correlated quantum phases are closely related to the presence of moiré flat bands (MFBs) [24-29].

One of the most representative moiré materials is twisted bilayer graphene, which features MFBs at discrete small angles (dubbed magic angles). Physically, the MFBs arise from the moiré pattern-induced interlayer hybridization, and can be explained by the magic angle mechanism [24-29]. Apart from the moiré superlattices formed with gapless graphene layers, it has been recently unveiled that ultraflat bands also exist in gapped moiré materials, such as twisted hexagonal boron nitrides and transition metal dichalcogenides [30-36], in which the formation of MFBs is understood in terms of the splitting of defect-like band edge states [33]. Interestingly, this mechanism does not require discrete magic angles, enabling much wider versatility than magic-angle twisted bilayer graphene.

Artificial crystals for various classical waves can provide exceptional experimental platforms for exploring the highly-intricate moiré physics, benefiting from the great flexibility in controlling lattice symmetries, interlayer couplings, and twist angles [37-50]. Some salient optical (acoustic) properties and phenomena have been observed in classical moiré structures [37-43], such as abnormal light localization and delocalization [37], enhanced nonlinearity and lasing effects [38-40], and moiré-fringe induced gauge fields [41,42]. Meanwhile, the magic angle mechanism for the formation of MFBs has been theoretically extended to the classical moiré materials [46-50]. However, to the best of our knowledge, the MFBs in classical moiré structures have not yet been experimentally observed in frequency-momentum spectra. The key challenges for directly observing MFBs lie in demanding a huge (defectless) sample at a (sensitive) small twist angle and resolving the MFBs in an extremely narrow frequency-momentum window. In fact, for similar reasons, direct energy-momentum resolved measurements remain elusive in quantum moiré materials [51,52], where the presence of MFBs is mostly inferred from indirect transport signals and scanning tunnelling spectroscopies [35,36].

In this Letter, we design a square-lattice moiré sonic crystal (SC) by twisting two gapped constituent monolayers at a large angle ($\simeq 22.6°$), and present a direct experimental evidence for the presence of isolated in-gap MFBs with frequency-momentum spectra. Moreover, we identify the localized nature of the MFB states by scanning their spatial distributions.

Unlike the MFBs in twisted bilayer graphene, which emerge for twist angle close to magic conditions, here ultraflat bands can be formed over a broad range of twist angles. Our findings have important implications for the future study of MFB physics in various moiré structures that modulate quantum or classical waves.

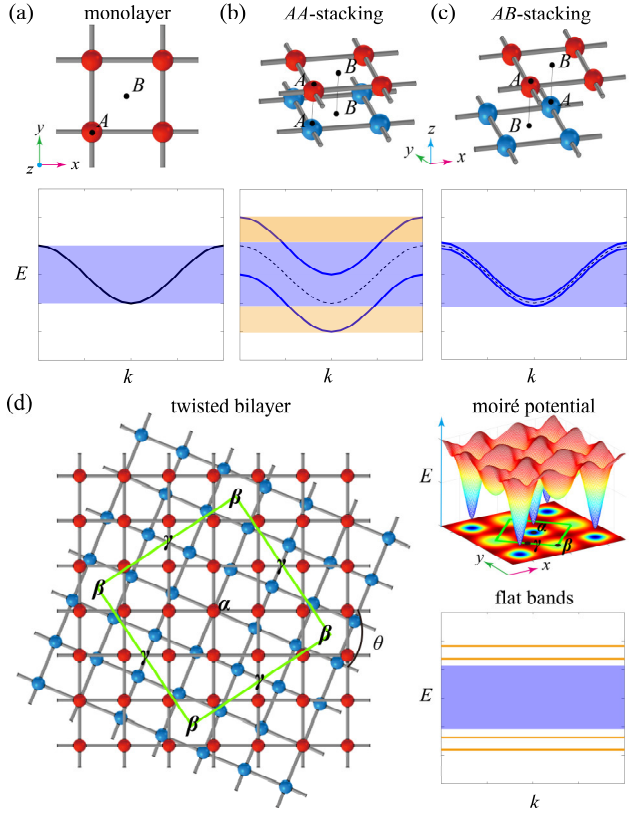

FIG. 1. Schematics for the formation of MFBs. (a)-(c) Lattice structures (top) and dispersion curves (bottom) for the monolayer, *AA*-stacked bilayer, and *AB*-stacked bilayer systems, respectively. (d) Twisted bilayer system. The green square denotes a moiré supercell, where $\alpha$ ($\beta$) and $\gamma$ label representative spatial domains of different orbital misalignments, resembling the *AA*- and *AB*-stackings, respectively. Ultraflat bands are formed by the energy states localized at the moiré potential wells ($\alpha$ or $\beta$) surrounded by the barriers ($\gamma$).

*Mechanism for the MFBs in gapped systems.*—We start with a simple square-lattice monolayer, which hosts a band well separated from the others [Fig. 1(a)]. The energy band splits when a pair of such monolayers are stacked, where the degree of band repulsion depends on the strength of interlayer coupling. Typical bilayer configurations are dubbed *AA*-stacking [Fig. 1(b)] and *AB*-stacking [Fig. 1(c)], where *A* and *B* represent the two high-symmetry points of the square lattice. In general, these high-symmetry stackings, which exhibit minimal and maximal interlayer misalignments, maximize and minimize the energy splitting, respectively, because of the extreme spatial overlaps of Bloch wavefunctions. Once the two layers are twisted, the spatially-varied misalignment forms a moiré superlattice [Fig. 1(d)]. The superlattice can host energy states localized at specific spatial domains (i.e., moiré potential wells), which hybridize to form flat bands by weak tunneling among the wells. Physically, the tightly confined states stem from the potential landscape induced by spatially-varied interlayer couplings [30]. More specifically, the energy states preferentially residing at $\alpha$ (or nearly equivalent $\beta$) regions, resembling those Bloch states in *AA*-stacking [Fig. 1(b), yellow shaded], will be confined when falling into the band gap of their surrounding $\gamma$ regions that resemble *AB*-stacking locally. With the reduction of $\theta$, both the size and distance of the potential wells increase. The former enables more states localized in individual potential wells, which evolve into more MFBs through intercell couplings; the latter weakens the intercell coupling and flattens the MFBs further. Obviously, this mechanism is applicable to any 2D lattice.

*Constructing moiré SCs with ultraflat bands.*—The above idea can realized with SCs. As shown in Fig. 2(a1), our monolayer SC consists of a square array (of lattice constant $a = 16$ mm) of identical air cavities (of size 14 mm × 14 mm × 8 mm) connected by square tubes (of side length 2 mm). Physically, the air cavities emulate atomic orbitals and the narrow tubes introduce couplings between them. As shown in Fig. 2(b1), the monolayer structure is designed to host a wide band gap between the $S$ and $P_{x(y)}$ bands developed respectively from monopole and dipole cavity modes. As sketched in Figs. 2(a2) and 2(a3), bilayer systems can be constructed by introducing a vertical disk (of radius 5 mm and height 2 mm) for each cavity, where the interlayer couplings depend mostly on the contact areas of the top and bottom disks. Particularly, here the disk radius is optimized to obtain a strong interlayer coupling in *AA*-stacking and a zero coupling in *AB*-stacking. Benefiting from the sharp contrast between these two cases, as shown in Figs. 2(b2) and 2(b3), the split $S_2$ band in *AA*-stacking *entirely* falls into the band gap of the *AB*-stacking sample (without energy splitting in the bands because of the vanishing interlayer coupling).

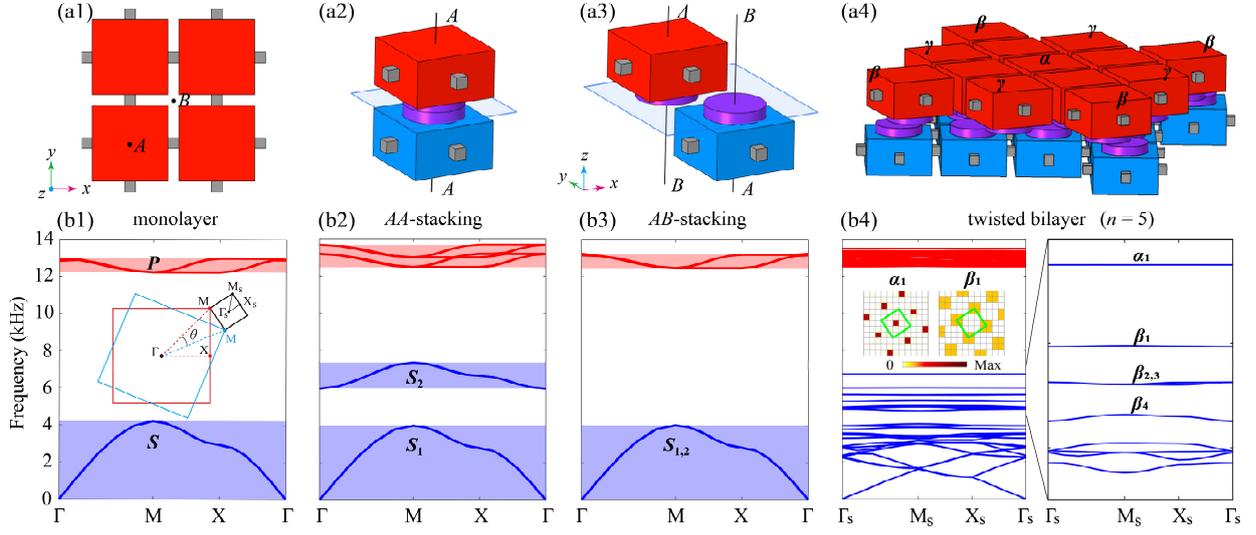

FIG. 2. Moiré SC and its flat bands. (a1)-(a4): Geometric structures of the monolayer, *AA*-stacked bilayer, *AB*-stacked bilayer, and moiré SCs, respectively. (b1)-(b4): The corresponding band structures. The inset in (b1) shows the first Brillouin zone of the square lattice as well as the resultant moiré Brillouin zone for a commensurate twist angle 22.6°. The insets in (b4) display the eigenstate distributions of the isolated MFBs $\alpha_1$ and $\beta_1$, where each grid represents a cavity on the top layer of the moiré SC, and the green square indicates a moiré supercell.

Similar to Fig. 1(d), we construct a moiré SC by rotating the bottom layer at an angle $\theta$ [Fig. 2(a4)]. Again, the central and corner regions of the moiré supercell are denoted with $\alpha$ and $\beta$, respectively, separated by the (fully gapped) $\gamma$ regions. For definiteness, we consider commensurate angle $\theta = 2\arctan(m/n)$, with $m$ and $n$ being coprime integers, and the moiré period reads $L = \sqrt{n^2 + m^2}a$ or $L = \sqrt{(n^2 + m^2)/2}\,a$ for an odd or even value of $|n-m|$. Here we take $m = 1$ and an odd value of $n$ to reduce the size of supercell. Figure 2(a4) exemplifies a moiré SC with $n = 5$, featuring a large twist angle $\theta \simeq 22.6°$. Its band structure [Fig. 2(b4)] shows clearly the emergence of multiple flat bands inside the band gap of the *AB*-stacked system. For convenience, we label the MFBs $\alpha_i$ and $\beta_i$ according to the spatial domains where the states are localized, with the subscript $i$ ordering the MFBs from high frequency to low frequency. In particular, the flattest band $\alpha_1$ appears at a frequency closer to the gap center. It remains flat even for the largest commensurate angle $\theta \simeq 36.9°$ (with $n = 3$). The inset of Fig. 2(b4) exemplifies the pressure intensity distributions for the eigenstates $\alpha_1$ and $\beta_1$ at $M_S$-point, which exhibit spatially confined moiré patterns as expected. The MFBs arising from the split $P_{x(y)}$ bands, less obvious due to the narrower frequency window exhibited in the band structures of the high-symmetry stackings, are not specified here.

A systematic study was performed for the moiré SCs (see *Supplemental Materials*). As expected, more MFBs appear as the increase (decrease) of $n$ ($\theta$), and more importantly, the moiré bands are flattened quickly. Note that the band flatness can be captured by an average group velocity of the Bloch states over the moiré Brillouin zone, which can be estimated as $\bar{v}_g \simeq$ $2L\Delta f$ with $\Delta f$ being the frequency bandwidth. This definition deducts the superficial flattening effect induced by pure band folding. As shown in Fig. 3(a), the average group velocity $\bar{v}_g$ becomes vanishingly small with respect to the sound speed in free air space, $c_0 = 340$ m/s, especially for the predominant MFB $\alpha_1$ ($\bar{v}_g/c_0 \simeq 0.1\%$ even for the case of $n = 3$). Almost, $\bar{v}_g$ diminishes exponentially with $n$ (or equivalently, with the moiré period $L \simeq na/\sqrt{2}$). This can be explained by the exponentially decayed coupling (roughly, $\propto \Delta f$) between the moiré supercells, $J \sim e^{-d/l}$, where $l$ characterizes the decay length of an in-gap state, and $d$ (roughly, $\propto L$) is the spatial distance between the localized states. This is markedly different from the magic angle mechanism where the flatness oscillates with $\theta$. Besides, the decay length $l$ decreases as the band moves to the gap center, which explains qualitatively the flattest MFB $\alpha_1$, as well as the fact that a strong interlayer coupling in *AA*-stacking (enforcing a strong band repulsion) favors the emergence of ultraflat bands (see *Supplemental Materials*). Furthermore, to quantitatively characterize the localization degree of the MFB states, we introduce a size-relevant inverse participation ratio (IPR), $\chi = N \sum_{i=1}^{N} I_i^2 / (\sum_{i=1}^{N} I_i)^2$, where $I_i$ denotes the pressure intensity of the $i^{\text{th}}$ cavity and $N = (n^2 + 1)/2$ is the site number of a moiré supercell. Comparing with the conventional IPR definition [53,54], here an extra factor $N$ is introduced to cancel the size effect in periodic systems. Physically, the larger is the value of $\chi$, the stronger is the mode localization. As shown in Fig. 3(b), indeed, the MFB states become more and more localized with reduced $\theta$, consistent with the dramatically flattened MFBs [Fig. 3(a)].

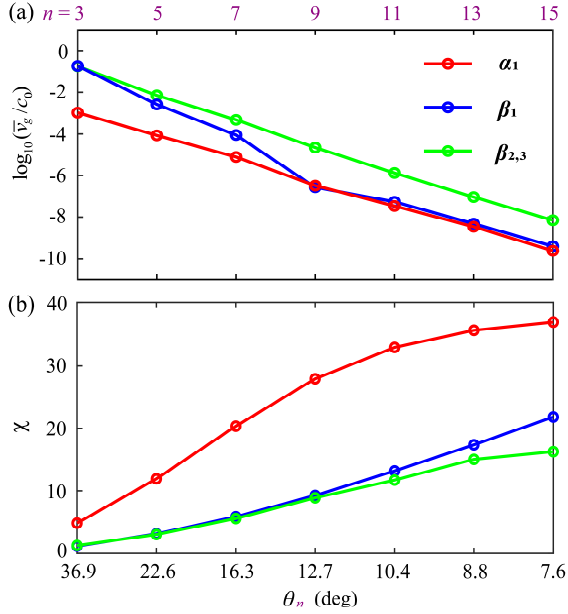

FIG. 3. (a) Normalized average group velocities ($\bar{v}_g$) and (b) size-relevant IPRs ($\chi$) of the first four MFBs ($\alpha_1$, $\beta_1$, and the nearly degenerate $\beta_2$ and $\beta_3$) for different commensurate angles, $\theta_n = 2\arctan(1/n)$.

*Experimental evidence for acoustic MFBs.*—The MFBs and associated localization states were identified unambiguously by our airborne sound experiments. Based on 3D printing technique, we fabricated a moiré SC of twist angle $\theta \simeq 22.6°$ and moiré period of $L \simeq 57.7$ mm. The sample contains $9 \times 9$ supercells (2106 cavities in total). In addition, a monolayer SC and an *AA*-stacked bilayer SC were fabricated for providing comparative band structures, both of which consist of $15 \times 15$ unit cells. The *AB*-stacked sample, essentially formed by two uncoupled monolayers, was not considered here.

Figure 4(a) shows our experimental setup, exemplified for the moiré sample (see *Supplemental Materials*). To detect desired sound responses, small holes were perforated on the (top-layer) air cavities for inserting the point-like sound source and detector, which were blocked when not in use. Both the input and output signals were recorded and frequency-resolved with a multi-analyzer system (B&K Type 3560B). Prior to presenting a systematic experimental study on the moiré SC, we measured the band structures for the monolayer and *AA*-stacked bilayer samples. To do so, we placed the sound source in the middle of the sample and scanned the pressure response cavity by cavity. Figure 4(b) shows the 2D spatial Fourier spectra (color scale) performed for the measured pressure fields. Comparing with the monolayer case, the band structure measured for the *AA*-stacked sample demonstrates clearly a dispersive band (of width ~1.4 kHz) falling inside the wide band gap. The experimental data reproduce well our numerical results (black dashed lines). The band broadening is mainly caused by finite-size effect and unavoidable acoustic dissipation.

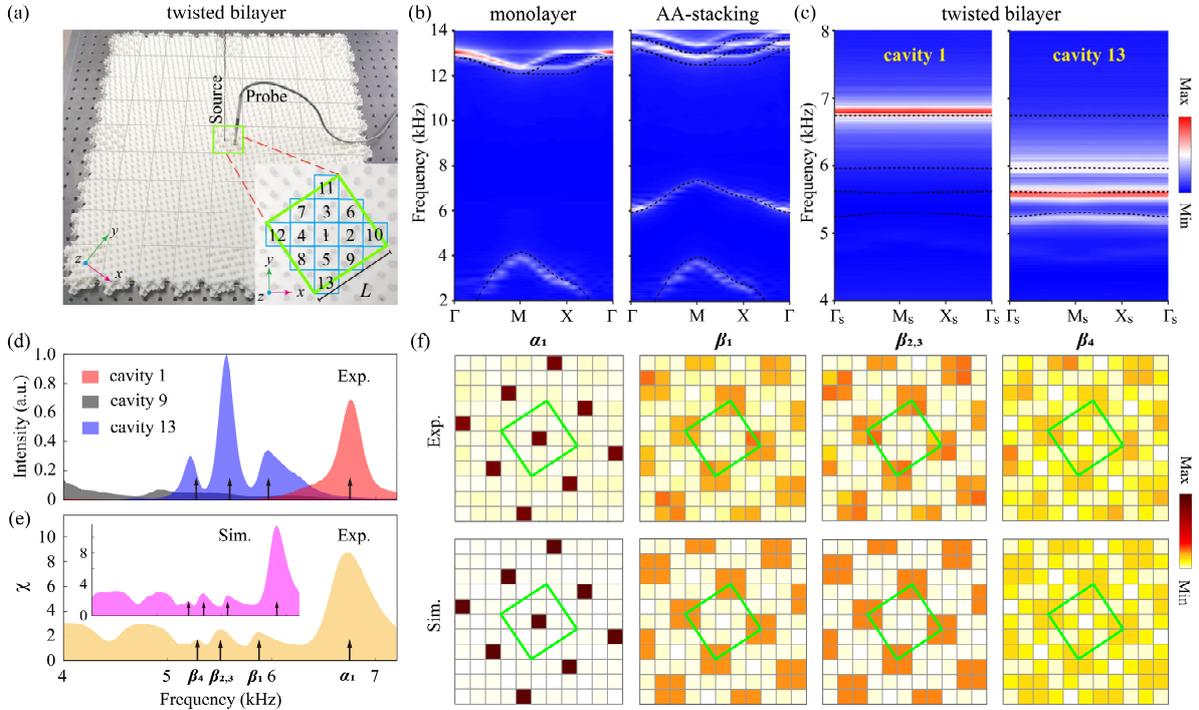

FIG. 4. Experimental observation of the MFBs and the associated sound localization. (a) Experimental setup. The inset shows a moiré supercell (green square) and its 13 top-layer cavities (blue squares) numbered for experimental detection. (b) Experimentally measured

Fourier spectra for our monolayer and *AA*-stacked bilayer SCs (color scale), comparing with the simulated band structures (black dashed lines). (c) Fourier spectra measured for the twisted-bilayer sample. The left and right panels correspond to different excitation-detection configurations. (d) On-site intensity spectra detected for three different cavities. The black arrows indicate the simulated MFB frequencies. (e) Measured IPR spectrum. Inset: numerical comparison. (f) Measured (top) and simulated (bottom) intensity distributions at the peak frequencies of (e).

For the twisted bilayer sample, we positioned the sound source and detector in specific cavities to excite desired MFBs, given the strong spatial preference of the localized eigenstates [Fig. 2(b4)]. The preference was confirmed first by a simple on-site measurement, where the sound source and detector were located at the same cavity of the middle supercell of the sample. Typically, the cavities 1, 13, and 9 were considered [Fig. 4(a), inset], which belong to the $\alpha$, $\beta$, and $\gamma$ regions of the supercell, respectively. Figure 4(d) presents the measured pressure intensity spectra. As expected, the spectrum for cavity 1 shows a predominant peak that corresponds to the MFB $\alpha_1$, and the spectrum for cavity 13 exhibits three peaks near the MFBs $\beta_1$, $\beta_{2,3}$, and $\beta_4$, in contrast to the negligible sound signal detected at cavity 9. Figure 4(c) presents the measured Fourier spectra for the cavities 1 and 13; in each case, the pressure information of all periodically equivalent cavities was scanned, given the sound source fixed at the cavity 1 or 13 of the middle supercell. Unambiguously, the experimental data capture the numerical MFBs selectively. Particularly, the measured MFBs are well-resolved in frequency, benefiting from the wide band gap in our ingenious acoustic design. Furthermore, we performed on-site measurements over multiple supercells to identify the moiré patterns inherited in the localized states. Figure 4(e) shows the measured IPR spectrum, which exhibits four peaks associated to the MFBs $\alpha_1$, $\beta_1$, $\beta_{2,3}$, and $\beta_4$, consistent with our numerical result (inset). Figure 4(f) presents the directly mapped intensity distributions at the four peak frequencies (first row), again in good agreement with the simulations (second row).

*Discussions and conclusions.*—We successfully observed the MFBs and associated localized states in an acoustic moiré structure. The combination of the wide band gap and big twist angle (thus allowing sufficient supercells in an acceptable sample size) enables excellent resolution of the MFB in frequency-momentum space. Although we focus on a less studied square lattice [55], the underlying physics, markedly different from the established magic angle mechanism, preserves in any moiré superlattice (see Supplemental Materials). Our findings can be easily extended to other classical artificial structures and promote promising application scenarios for controlling waves [37-43]. For example, the high density of states of the MFBs can be used for enhancing light (sound) trapping, lasing, and nonlinearity, the moiré patterned optical (acoustic) fields enable exotic particle manipulations, and the drastically reduced group velocity would also be useful in on-chip light (sound) signal processing. Last but not the least, the conclusive evidence for the new MFB mechanism would be enlightening to electronic systems [30-36] and impel the applications of moiré materials to practical nanometer integrated devices (benefited again from the big twist angles and small samples).

**Acknowledgements**
This project is supported by the National Natural Science Foundation of China (Grant No. 11890701, 12004287, 12104346), the Young Top-Notch Talent for Ten Thousand Talent Program (2019-2022), the National Postdoctoral Program for Innovative Talents (Grant No. BX20200258), and the China Postdoctoral Science Foundation (Grant No. 2020M680107).

**Contributions**
C.Q. conceived and supervised the project. X.Z. and T.L. performed numerical simulations. T.L. and X.Z. designed the samples and did the experiments under the assistance of Q.Z. and X.F. C.Q. wrote the manuscript with input from all authors.